# Channel Sounding for the Masses: Low Complexity GNU 802.11b Channel Impulse Response Estimation

Mohammad H. Firooz, Dustin Maas, Junxing Zhang, Neal Patwari, and Sneha K. Kasera

**Abstract**

New techniques in cross-layer wireless networks are building demand for *ubiquitous channel sounding*, that is, the capability to measure channel impulse response (CIR) with any standard wireless network and node. Towards that goal, we present a software-defined IEEE 802.11b receiver and CIR estimation system with little additional computational complexity compared to 802.11b reception alone. The system implementation, using the universal software radio peripheral (USRP) and GNU Radio, is described and compared to previous work. By overcoming computational limitations and performing direct-sequence spread-spectrum (DS-SS) matched filtering on the USRP, we enable high-quality yet inexpensive CIR estimation. We validate the channel sounder and present a drive test campaign which measures hundreds of channels between WiFi access points and an in-vehicle receiver in urban and suburban areas.

## I. Introduction

Channel impulse response (CIR) measurements have long held importance for communication system design [1], [2], [3], [4], [5]. The CIR describes the spreading, or echoing, that occurs when an impulse is sent through a channel. This spreading in time can lead to inter-symbol interference (ISI), and frequency-selective or narrow band fading, depending on the symbol bandwidth. A knowledge of the CIR characteristics enables system designers to ensure that ISI does not dominate and hence lead to an excessive irreducible bit error rate [6]. Multipath channels can also be used to increase the bit rate and reliability of multiple-input multiple-output (MIMO) communications systems. Accurate MIMO channel models can be built from CIR measurements [7], and can be used to improve MIMO system design [8]. In general, measurements of CIR in wireless networks have become increasingly important to determine the real-world performance of many new technologies.

M.H. Firooz is with the Dept. of Electrical Engineering, University of Washington, Seattle, USA. D. Mass and N. Patwari are with the Department of Electrical and Computer Engineering, University of Utah, Salt Lake City, USA. J. Zhang and S.K. Kasera are with the School of Computing, University of Utah, Salt Lake City, USA. This material is based upon work supported by the National Science Foundation under Grant Nos. #0855261 and #0831490, and a University of Utah Research Foundation TCP Grant. Correspondance email: npatwari@ece.utah.edu.

Several new cross-layer wireless networking technologies use measurements of the multipath channel for purposes of environmental awareness and security. In fact, the capability of standard wireless devices to measure CIR can enable many new applications and security features for wireless networks. For example:

- *Improved Ranging*: Devices that use Time of Arrival (TOA) or Time Difference of Arrival (TDOA) for localization can measure CIR and use CIR statistics to determine whether a link was line-of-sight (LOS) or non-line-of-sight (NLOS) [9], and thus whether its range measurement was reliable or unreliable.
- *Fingerprint-based Localization*: Measured CIR can be used as a fingerprint of the location of a device [10]. Measurements can be made during a calibration phase at each position; later, a device's measured CIR can be compared to the calibration database to locate it. Measurements could even be partly replaced by ray-tracing simulations; in this case, it will be critical to accurately compare ray-tracing predictions and CIR measurements [11].
- *RF-based Multistatic Radar*: Through-wall surveillance (TWS) is possible at RF using measured CIR of a radar channel. Today's TWS devices use ultra-wideband transceivers, but research is also investigating narrowband versions [12], [13]. It is possible that a deployed wireless communication network can be used for passive imaging within buildings [14], [15].
- *Location Distinction*: For security reasons, we may be interested in knowing solely if a device has *changed* position, even if we cannot estimate its position. Location distinction provides a type of location-based authentication using measurements of the CIR over time [16], [17]. When measured CIRs are consistent with past measurements, the network knows that the user is not an impersonator and is stationary. Such information is useful in both network security and physical security of radio-tagged objects that are supposed to be stationary.
- *Secret Key Establishment*: The CIR on a link varies unpredictably as a function of the two node positions, and is largely reciprocal in both directions of the link. Thus it can be used to generate a secret key that is known only to the two nodes [18], [19], [20], [21]. Measurements of the CIR over time can be encoded into a bit key that is very likely to agree at both ends of the link [22].

These applications require CIR measurements to be performed in real time using commercial wireless devices, as opposed to with specialized measurement equipment or in post-processing. Typical commercial wireless devices use the received signal in a demodulator to estimate the transmitted bits, but then discard the received signal samples. Information about the channel (besides the received signal strength) is not forwarded to higher networking layers, nor can it be estimated from the demodulated bits. For the above applications to be viable, future commercial wireless devices must be able to rapidly calculate impulse response information. At present, in order to enable measurement-based research for these applications, we require hardware that does measure and record CIR during each packet reception. In this paper, we present the design of an inexpensive open-source CIR measurement system that enables researchers to quickly and accurately perform large-scale CIR measurements. Using any commercial 802.11b transmitter, it is possible to sound the channel between that transmitter (separated by its MAC address) and our system, without any modification at the transmitter side.

We build our system on commercially available Universal Software Radio Peripherals (USRPs). Compared to expensive signal analyzers and oscilloscopes, our system is low cost. Moreover, one can purchase USRPs and deploy our system in large networks as might be seen in typical WiFi deployments. Our system works seamlessly with standard PHY layer signals from commercial wireless devices. To demonstrate CIR measurement-based applications when using the standard bandwidth and standard modulations that will exist in real-world systems, we provide a low-cost open-source system implementation that measures CIR and demodulates standard 802.11b signals. Our implementation, available online at [23], provides, essentially, an 802.11b receiver with the additional capability of CIR estimation. Specifically, we provide an implementation of an 802.11b FPGA matched filtering method, the first, to our knowledge, to be presented for the USRP-based GNU radio framework. We also provide a method to estimate the CIR from a modulated 802.11b signal. In particular, we use the output of the receiver's matched filter, which allows a low-complexity CIR estimate. Furthermore, we apply a deconvolution algorithm by which estimated multipath temporal delays to the nearest 30 ns. Finally, the accuracy of the system is verified by measuring test channels in the lab.

Computation of a CIR from 802.11g OFDM signals requires more computational resources than from 802.11b DS-SS signals. However, this work provides the capability to perform experimentation with 802.11b, which is sufficient to characterize the performance of the applications listed above.

Our implementation shows that accurate CIR estimation can be performed using a resource-constrained FPGA, which provides a proof-of-concept for future commercial devices. We also perform measurements using known channels to verify channel accuracy. Using our implementation, we perform a drive test, recording CIRs in six different areas in Salt Lake City. We find that our measurement system produces expected results. Our results show that in residential areas, there are few multipath at significant delays, while in commercial and downtown areas, there are multipath at longer delays, and with amplitudes close to the amplitude of the first arriving path. Furthermore, average RMS delay spreads, that are averages over all measured channels ($\approx$ 3 million measurements), range from 10.8 ns in the lowest multipath (residential), to 68.9 ns in the highest multipath (downtown) area.

The rest of this paper is organized as follows. The next two sections present a background on CIR measurements and describe our system while highlighting how our work differs from existing work. Section IV presents the methods used in this paper for CIR estimation. The implementation of these methods is described in Section V. Section VI provides an experimental verification and results of a measurement campaign. Finally, Section VII concludes.

## II. Background

In general, CIR measurements are performed with specialized channel-sounding measurement equipment. These measurement methods can be categorized as follows [24]:

- *Pulse Envelope Measurement Systems*: A transmitter sends a short duration pulse as an approximation of an

impulse function, and a receiver measures the amplitude of the received signal [1], [25]. A large, interference-free bandwidth is required.

- *PN Sequence Generator / Sliding Correlator System*: A pseudo-noise (PN) signal is generated and continuously transmitted. A receiver correlates the received signal with a copy of the known PN signal. The receiver PN signal rate is slightly slower, which allows the correlator output to be recorded at the difference frequency, a much slower rate. First proposed in [2], this method has been used many times in the past few decades with a variety of bandwidths, center frequencies, radio environments, and link lengths, including in [5], [26], [27], [28], [29], [30]. An overview of sliding correlator systems has recently been published in [31]. The hardware is specialized and not available from standard RF test equipment vendors – such a transmitter and receiver are typically built from parts.
- *Frequency Domain Channel Measurement*: In these systems, a vector network analyzer is used to measure the frequency response of a channel [32], [33], [34]. Such systems are limited in range because the transmit antenna and receive antenna must both be connected by cable to the network analyzer.

The above channel sounding methods involve measurement systems set up for the sole purpose of channel impulse response measurement, and as such, are designed to achieve high performance (in terms of temporal resolution and dynamic range). However, these are not communication systems – none of the systems communicate information from the transmitter to the receiver. They require a wide bandwidth, which would not necessarily be available for commercial use. Our research requires no bandwidth utilization, in the sense that we record CIRs from ambient 802.11b signals transmitted for other purposes. In addition, the described hardware is expensive; for example, a 3-GHz vector network analyzer can cost US $20,000. These channel sounding methods were not designed for implementation in standard wireless communication devices. Our research in this paper is motivated by these limitations.

## III. Our System

The channel sounder we present in this paper for 802.11b is similar to the existing sliding correlator channel sounding method, in that it uses the fact that a PN sequence is repeatedly transmitted by 802.11b transmitters. However, it is different from the existing method in the following four significant ways:

1) The PN sequence in 802.11b is fixed to be the 11-chip Barker sequence, which is not designed for high dynamic range CIR estimation.
2) Devices transmitting in 802.11b send PN-coded symbols modulated with data; modulation is undesirable from the perspective of CIR estimation.
3) The implemented receiver is not a “sliding correlator”, instead it performs correlation of the received signal in real time using the same rate PN signal as the transmitter.
4) The channel between the channel sounder and any standard 802.11b transmitter (*e.g.*, laptop or access point) is measured; no specialized transmitter is required. Since WiFi devices are located throughout buildings, homes, and cities, there are always many channels that can be rapidly measured.

Note that IEEE 802.11b devices must support two mandatory bit rates (1 Mbps and 2 Mbps) and may optionally support two higher rates (5.5 Mbps and 11 Mbps) as specified in [35]. In this work, for simplicity, we only consider the standard rates. We note that the start of any 802.11b packet and some 802.11g packets (the first 192 symbols, known as the PLCP), are sent at either the 1 or 2 Mbps rate. Thus many sources exist that our system implementation can utilize for CIR estimation.

The system presented is comprised of open source software and hardware. It is built upon GNU Radio, which was started in 1998 as an open source framework for software-defined radio [36][37][38]. The system uses the universal software radio peripheral (USRP), an open-source transceiver platform. Its receiver hardware includes omnidirectional antenna, (swappable) downconverter, analog-to-digital converter (ADC), and field-programmable gate array (FPGA) [39]. The USRP is the only 'cost' of the proposed system, and we mention that its current price is US$975.

The USRP is often used to sample a radio signal, filter it to the signal bandwidth, and then transfer the samples to a PC for software demodulation and processing. However, USRP-based software radios are limited in their ability to receive 802.11b signals in this manner due to a bandwidth bottleneck in the transfer of data from the USRP to the PC through a USB 2.0 link. Effectively, as discussed in Section V, a USB 2.0 link can only transfer eight million complex samples per second, or 16 million real-valued samples per second (16 MS/s) at eight bits per sample. In contrast, the Nyquist sampling rate of an 802.11b signal is 22 MS/s. Two solutions to the bandwidth bottleneck are compared and contrasted in this paper:

1) *Bandwidth Reduction Method*: Previous work filters 802.11b signals from 22 MHz down to 8 MHz RF bandwidth, prior to sampling [40][38]. The filtered signal requires a 16 MS/s sampling rate, which can then be transferred over the USB to the PC. Filtering the signal to 8 MHz (RF bandwidth) severely degrades the received signal, which effectively has a very low SNR. Bandwidth reduction both reduces the performance of the receiver and degrades the temporal resolution of any CIRs estimated from the received signal. This implementation is the state-of-the-art 802.11b receiver developed for GNU radio.
2) *FPGA Matched Filtering Method*: In this method, the received 802.11b signal is despread on the FPGA prior to sampling. Despreading an 802.11b signal reduces the sample rate (as low as 2 MB/s) that must be sent through the USB link, without degrading the received signal. It also reduces the computational complexity of the PC-based receiver software. One major contribution of this paper is to show how this despreading can be accomplished within the strict computational limitations of the FPGA used in the USRP.

In a closely related project, a channel sounder for 802.11b applications is reported by Jemai and Kürner in [41]. This channel sounder first records the samples of a 192-bit segment of the 802.11b signal onto a PC. Then, the signal is despread and demodulated on the PC. Next, the transmitted signal for the 192-bit segment is recreated using the demodulated bits. Finally, the recorded received signal and recreated transmitted signal are convolved. Since both have many samples, the cross-correlation consumes significant PC computation time, on the order of $NB \log NB$, where $N$ is the number of samples per bit, and $B$ is the number of bits used. In comparison, our system involves PC computation on the order of $NB$. The system proposed in [41] uses proprietary software and VHDL implementations (ComBlock products from Mobile Satellite Services Inc.), while our implementation uses open-source hardware and software (GNU Radio) with a wide user base that utilizes and contributes to the code library.

Our specific contributions to 802.11b CIR estimation system research are summarized as follows:

1) We provide an implementation of an 802.11b FPGA matched filtering method, the first, to our knowledge, to be presented for the USRP-based GNU radio framework.
2) We provide a method to estimate the CIR from a modulated 802.11b signal. In particular, we use the output of the receiver's matched filter, which allows a lower-complexity CIR estimate compared to [41].
3) We describe a deconvolution algorithm by which we improve the temporal resolution of our CIR estimates to 30 ns.

We perform extensive tests to verify that the platform works correctly, in both lab-controlled and real-world multipath channels.

## IV. Analysis Methods

In this section, we present a detailed analytical framework for CIR estimation using received 802.11b signals. The WiFi 802.11b PHY layer is a direct-sequence spread-spectrum (DS-SS) signal, that is, it is a signal spread across a wide bandwidth in order to achieve particular desirable properties. We describe the CIR estimation process within this particular PHY layer. It is possible to estimate the CIR from signals from other protocols, however, these are not the focus of this paper. For example, it is possible to estimate the CIR from an 802.11g signal, which uses an orthogonal frequency division multiplexing (OFDM) PHY layer signal. We note that 802.11g devices may also transmit 802.11b packets, and in fact, often operate in an DSSS-OFDM mode in which the physical layer convergence protocol (PLCP) preamble and header are transmitted using DS-SS and the remaining data sent using OFDM. It is possible to estimate CIR only from the DS-SS symbols that comprise the PLCP preamble and header. If the PLCP is known a priori, then reception range can be significantly greater by using the whole PLCP packet; Because the "energy per bit" is essentially increased by a factor of 48, essentially a 17 dB increase. However, since we want a channel sounder that works with any 802.11b transmitter, we cannot know the PLCP ahead of time.

In short, we present in this section how an 802.11b signal is described, how it is impacted by a multipath channel, and how the proposed system estimates both: (1) the transmitted data, and (2) the amplitudes and delays of the multipath in the channel. The presented 802.11b signal framework is used throughout this paper.

### A. Transmitted Signal

The 802.11b physical layer uses direct-sequence spread-spectrum (DS-SS) modulation. Each transmitted symbol has duration $T_s = 1\mu$s, and the $j$th transmitted data symbol is denoted $b_j$. In DS-SS, this transmitted symbol stream is multiplied by a pseudo-noise (PN) code signal, which also has duration $T_s$. Denoting the PN code signal as $c(t)$, the transmitted signal in baseband is given by

$$s(t) = \sum_j b_j c(t - jT_s).$$

Note that $b_j$ generally takes complex values, because data symbols may be modulated either using differential binary phase-shift keying (DBPSK) or differential quadrature phase-shift keying (DQPSK).

The PN code in 802.11b is called the Barker code. This code consists of eleven *chips*, each with duration $T_c = T_s/11$ $\mu$s. Chips have an amplitude of either 1 or -1, and denoting the $i$th chip value as $c_i$, the Barker code is the chip sequence

$$\mathbf{c} = [c_0, \ldots, c_{10}]^T = [+1, -1, +1, +1, -1, +1, +1, +1, -1, -1, -1]^T \tag{1}$$

The Barker code signal is a modulated sequence of pulses with amplitude given in (1). Specifically,

$$c(t) = \sum_{i=0}^{10} c_i p(t - iT_c), \tag{2}$$

where $p(t)$ is the pulse shape. The pulse shape is chosen to meet the bandwidth limitations imposed by the 802.11b standard, but the precise shape of $p(t)$ is left to the designer. In this paper, when it is necessary to use a particular pulse shape, we have chosen to use a square root raised cosine (SRRC) pulse with roll-off factor $\alpha = 0.5$, which meets the spectral mask requirements and represents a good tradeoff between temporal and frequency domain characteristics [42].

The PN code signal has a high bandwidth because it switches chips at a high rate, specifically, eleven times the symbol rate. The bandwidth of the transmitted data signal is "spread" to eleven times the bandwidth of the original data symbol signal. For 802.11b, this transmitted DS-SS signal has a chip rate of 11 MHz. Hence, transmitted 802.11b signals consume a wide bandwidth compared to the symbol rate of 1 MHz. As we will show, the wide bandwidth benefits CIR estimation by allowing higher temporal resolution; but it will also require despreading in the receiver, which can be a computationally expensive operation.

### *B. Multipath Channel*

Because of the multipath radio channel, many copies of the transmitted signal arrive at the receiver with different time delay, amplitude, and phase. Viewed as a filter, the multipath channel filter is a summation of impulses, each with the time delay and complex amplitude of a particular multipath,

$$h(t) = \sum_{l=0}^{L-1} \alpha_l \delta(t - \tau_l), \tag{3}$$

where $L$ is the total number of multipath, $\alpha_l = |\alpha_l| e^{\angle \alpha_l}$ is the complex amplitude gain of the $l$th multipath, $\tau_l$ is the delay of the $l$th multipath, and $\delta(\cdot)$ is the Dirac delta function. Since we are only interested in the relative time delay of each multipath, we let $\tau_0 = 0$, and then $\tau_l$ is the additional delay compared to the first arriving multipath.

### C. Reception

Since the channel is represented as a filter in (3), we can represent the received signal $r(t)$ as the convolution of the transmitted signal and the channel filter. Applying (3), we have

$$r(t) = s(t) \star h(t) = \sum_{l=0}^{L-1} \alpha_l s(t - \tau_l). \tag{4}$$

Where $s(t)$ is transmitted signal. Applying (1), we can write the received signal in terms of the symbols $\{b_j\}$,

$$r(t) = \sum_{l=0}^{L-1} \sum_j \alpha_l b_j c(t - \tau_l - jT_s).$$

In order to determine the symbols, an 802.11b receiver "de-spreads" the signal, that is, performs matched filtering with the PN code signal $c(t)$ from (2). A matched filter convolves the received signal with $c(-t)$ and samples the result once per symbol. We denote this signal as $q(t) = r(t) \star c(-t)$. We also define $R_c(t) = c(t) \star c(-t)$, so that we have

$$q(t) = \sum_{l=0}^{L-1} \alpha_l \sum_j b_j R_c(t - \tau_l - jT_s). \tag{5}$$

Note that $R_c(0)$ is the energy in the signal $c(t)$, which we denote $\mathcal{E}_c$.

It should be mentioned that in above formulation we do not consider interference effect from other transmitters, becuase in a DS-SS communication system, de-spreading process increase Signal to Interference plus Noise Ratio (SINR) by a factor which is the ratio of spread signal bandwidth to information signal bandwidth. In 802.11b this factor, usually denoted by $G$, is 11 [43]. Thus, after de-spreading we have eleven times reduction in interference. If the interference level is still too high, the packet will be dropped, because it will not pass the CRC check.

For the purposes of demodulation, we approximate $R_c(\tau)$ as zero for $\tau > 0$. As we will discuss in detail in Section V, this approximation is suitable for analysis of demodulation but unsuitable for analysis of CIR estimation. When $t = \tau_l + jT_s$, the value of $q(t)$ is then approximately equal to $\alpha_l b_j \mathcal{E}_c$. Since $\alpha_l$ is not a function of $j$, its phase is canceled in differential reception. Thus we can read the value of each symbol from the samples of $t = \tau_l + jT_s$, for $j = 0, 1, \ldots$. We could choose any $l$, but a good receiver would either try to select $l$ such that $|\alpha_l|$ was maximum, or use a rake receiver to use several multipath $l$ in the symbol decision.

The despreading operation is one that standard 802.11b receivers must perform in order to do symbol demodulation. We propose that $q(t)$ can be used directly in CIR estimation as well. By using an output that existing 802.11b receivers compute, we make it more feasible for future 802.11b receivers to estimate CIR without significant additional computational complexity.

### D. Correlation Signal: Data Effects

The estimation of channel impulse response from a received 802.11b signal is complicated by the data carried in the modulated data packet. To show these effects, we consider two cases of 802.11b signals:

1) *Ideal, Unmodulated Signal*: Although clearly unrealistic, we consider the case where the symbols transmitted in the 802.11b packet are identically equal to 1. That is, $b_j = 1$ for all $j$ and the PN code stream is unmodulated by symbol data. Specialized channel measurement systems such as sliding correlator systems (see Section I) use such signals to probe the channel.

2) *Realistic Modulated Signal*: The PN code signal is modulated with data, presumably unknown to the receiver until after demodulation. For example, for BPSK, $b_j \in \{-1, +1\}$.

Our comparison of these two cases in this section demonstrates the additional complexity added to CIR estimation by the use of 802.11b signals, compared to unmodulated DS-SS signals that would be possible using specialized measurement equipment.

First, we consider the effect of an ideal unmodulated signal in (5). The expression would simplify to

$$q(t) = \sum_{l=0}^{L-1} \alpha_l R_{pn}(t - \tau_l), \quad \text{where } R_{pn}(t) = \sum_j R_c(t - jT_s). \tag{6}$$

Here, $R_{pn}$ is the correlation of a PN code signal with a repeating PN code signal with period $T_s$. The Barker code has the property that this correlation function $R_{pn}(t)$ peaks at $t = 0$ and integer multiples of $T_s$ and is almost constant in between those peaks. This is a very desirable property for purposes of impulse response estimation and is a property of the 11-length Barker code [44]. Figure 1(a) shows the signal $q(t)$ when there is exactly $L = 1$ path with amplitude $\alpha_0 = 1$. The normalized amplitude of $q(t)$ is constant at $-1/11$ between the correlation peaks, which have amplitude 1. Multipath with delays of $\tau_l$ would add time-delayed versions of $q(t)$ to be summed together; the constant correlation in between peaks is desirable to identify these multipath contributions even when their magnitude $|\alpha_l|$ is small.

For realistic modulated signals, the correlation is not always constant in between the peaks, and low-amplitude multipath contributions would not be as easy to identify. In Figure 1(b), we use the transmitted symbols $\mathbf{b} = [1, 1, -1, 1, 1]$, and plot the correlation output signal $q(t)$ from (5), for the case that $L = 1$ and $\alpha_0 = 1$. Note that the normalized amplitude of $q(t)$ between the 2nd and 3rd peaks, and between the 3rd and 4th peaks, rapidly change between $\pm 1/11$. These periods of varying correlation correspond to the times in between changes in symbol values $b_j$. When $b_j \neq b_{j+1}$, the value of $q(t)$ for $jT_s < t < (j+1)T_s$ is not constant. But, note that when $b_j = b_{j+1}$, there is a nearly constant $-1/11$ correlation value in between the two peaks at $jT_s$ and $(j+1)T_s$. When subsequent symbols are identical, the constant correlation value in $q(t)$ can be exploited for improved CIR estimation.

### *E. CIR Estimation*

As seen above, we are motivated to use the correlator output signal $q(t)$ whenever the symbol value $b_j$ repeats. By doing so, we can avoid the negative impact of symbol modulation and still achieve high-performance CIR estimation. To formulate the effects, we now define two more correlation functions, $R_o(t)$ and $R_s(t)$, that represent the correlation output during symbol value changes, and during symbol value repeats, respectively,

$$R_o(t) = \begin{cases} R_c(t) - R_c(t - T_s), & 0 \le t \le T_s \\ 0, & o.w. \end{cases},$$

$$R_s(t) = \begin{cases} R_c(t) + R_c(t - T_s), & 0 \le t \le T_s \\ 0, & o.w. \end{cases}.$$

The functions $R_s(t)$ and $R_o(t)$ are shown in Figure 2 (a) and (b), respectively. We also define two subsets, $J_s = \{j : b_j = b_{j+1}\}$ for symbol integers $j$ when the next symbol value repeats, and $J_o = \{j : b_j \neq b_{j+1}\}$. Then we can write (5) as

$$q(t) = \sum_{j \in J_s} b_j \sum_{l=0}^{L-1} \alpha_l R_s(t - jT_s - \tau_l) + \sum_{j \in J_o} b_j \sum_{l=0}^{L-1} \alpha_l R_o(t - jT_s - \tau_l). \tag{7}$$

This version of $q(t)$ contains terms $R_s(\cdot)$ and $R_o(\cdot)$ that have support only over one symbol period.

From (7) we estimate the CIR by averaging only the symbol periods of $q(t)$ that correspond to repeated symbol values. This can be written,

$$\hat{h}(t) = \frac{1}{|J_s|} \sum_{j \in J_s} b_j q(t - jT_s) I_{(0,T_s)}(t) \tag{8}$$

where

$$I_{(a,b)}(t) = \begin{cases} 1, & a \leq t \leq b \\ 0, & o.w. \end{cases}.$$

Essentially, the channel estimator in (8) averages together only the impulse responses estimated during periods when the symbol values is not switching and thus the correlation function is constant.

$$\hat{h}(t) \approx \sum_{l=0}^{L-1} \alpha_l R_s(t - \tau_l) \tag{9}$$

Note that symbol values $b_j$ do not affect $\hat{h}(t)$. Effectively, the channel estimate is a sum of time-delayed, attenuated, and phase-shifted versions of $R_s(t)$.

### *F. Deconvolution*

Because the CIR estimate $\hat{h}(t)$ in (8) is a convolution of the true channel impulse response in (3) with $R_s(t)$, an improved estimate can be obtained via deconvolution. The function $R_s(t)$, shown in Figure 2, has a zero-to-zero pulse width of approximately 182 ns. Since multipath arrive more closely spaced than 182 ns, the complex-valued, time-delayed pulse shapes $R_s(t-\tau_l)$ overlap in time. It is difficult to visually inspect $\hat{h}(t)$ to identify when multipath arrive, unless subsequent multipath arrive separated by 182 ns. Such separation in delay implies additional path length of 61 m, which would be very rare in a short-range WiFi link – other multipath would likely exist in between two paths with such excess delay.

In this section, we describe the application of a deconvolution procedure which estimates the CIR with a 30 ns temporal resolution. This is not to say that any two multipath with 30 ns relative delay will be resolved all of the time – the limits of temporal estimators are generally determined by bandwidth, SNR, and signal duration [45], and are not explored in this paper. This estimator is optional; some applications would not need the additional resolution. We present the deconvolution-based CIR estimation algorithm here; its performance is validated via experimental tests in Section VI-B.

Other frequency-domain methods are possible for multipath delay estimation [11], [46], [47], but we apply the deconvolution method of [48] since the time domain signal $\hat{h}(t)$ is already available. Since $R_s(t)$ is known and largely constant for most of its period, deconvolution performs consistently and achieves a high dynamic range of multipath resolution.

It should be emphasized that traditional blind channel estimation algorithms, such as [49], recover the combined impulse response caused by the channel, transmit filter, and receiver filters. Since we want to separate the channel filter from the pulse-shaping effects, we cannot directly apply such blind channel estimation algorithms.

Although (9) is written in continuous time, the processing must deal with a discrete-time sampled signal as follows. Let

$$\hat{h}[n] \quad = \quad \sum_{l=0}^{L-1} \alpha_l R_s(n\, T_s - \tau_l) + w[k] \tag{10}$$

where $w[k]$ is measurement noise that is assumed to be i.i.d. Gaussian. It can be rewritten in vector form as,

$$\hat{\mathbf{h}} = \mathbf{R}_s \boldsymbol{\alpha} + \mathbf{w} \tag{11}$$

where $[\mathbf{R}_s]_{k,l} = R_s(kT_s - \tau_l)$ is an $M \times L$ matrix, $\boldsymbol{\alpha} = [\alpha_1, \ldots, \alpha_L]^T$ is an $L$-dimensional column vector containing the unknown weights $\alpha_l$, and $\mathbf{w}$ is $L$-dimensional noise vector. Equation (11) has $M$ equations ($M$ samples) and $L$ unknown parameters. The maximum likelihood estimate of $\boldsymbol{\alpha}$ is given by [48] as,

$$\min_{\boldsymbol{\alpha}} \|\boldsymbol{\alpha}\|_2, \text{ s.t. } \|\hat{\mathbf{h}} - \mathbf{R}_s \boldsymbol{\alpha}\|_2^2 \leq B \tag{12}$$

where $\|\cdot\|_2$ is the $l_2$-norm and $B$ is a fixed bound that constrains the sum of the squared residues. This can be written as the convex optimization problem,

$$\min_{\boldsymbol{\alpha}} \|\hat{\mathbf{h}} - \mathbf{R_s} \boldsymbol{\alpha}\|_2^2 + \lambda \|\boldsymbol{\alpha}\|_2^2 \tag{13}$$

where $\lambda$ is a fixed parameter that corresponds to the given $B$ from (12). In particular, $\lambda$ is the inverse of Lagrange multiplier of (12) at the optimum point [50]. The setting of $\lambda$ is discussed in [48]. The formula (13) can be rewritten in a way that leads to a quadratic programming problem:

$$\begin{aligned} \hat{\boldsymbol{\alpha}} \quad &= \quad \min_{\boldsymbol{\alpha}} \frac{1}{2} \boldsymbol{\alpha}^T \mathbf{H} \boldsymbol{\alpha} + \mathbf{f} \boldsymbol{\alpha} \\ \text{where } \mathbf{H} \quad &= \quad 2\mathbf{R}_s^T \mathbf{R}_s + \lambda \mathbf{I}, \text{ and } \mathbf{f} = -2\hat{\mathbf{h}}^T \mathbf{R_s} \end{aligned} \tag{14}$$

where $\mathbf{I}$ is the identity matrix. Vector $\boldsymbol{\alpha}$ is found by solving the quadratic programming problem in (14).

## V. Implementation

In this section, we present the system implementation of the WiFi-based receiver and CIR estimator. We have chosen to implement the system using open-source hardware and software as part of the GNU radio project. However, this paper is not limited in scope to the USRP. We intend to provide an implementation that provides practical CIR estimation in hardware with strict computational limitations. It will not be feasible to compute a CIR estimate with commercial 802.11 hardware unless the computation complexity is low, and the USRP provides a computationally-limited platform. The open-source platform is an advantage, we believe, because it is likely to lead to cooperative improvement in the system capabilities and large-scale adoption. Providing a system implementation that works within the limitations of the hardware platform is, in part, a demonstration of the feasibility of the approach in future commercial systems.

The capabilities and limitations of the USRP are as follows. The USRP receiver path consists of a 64 MS/s (million samples per second) 12-bit ADC, an Altera Cyclone FPGA, and a USB controller. The FPGA can be used to perform high rate filtering, but it is limited in the complexity of the operations that can be implemented.

In many software-defined radio (SDR) applications, the sampling rate of 64 MS/s is too high, and thus the FPGA is used solely to filter and downsample the sampled signal, and the USB controller is used to send data over a USB 2.0 link to a PC. However, the bandwidth of a USB 2.0 link, using a high-quality USB controller, is limited to approximately 32 MB/s [37] which is not sufficient to stream 802.11b signal samples. The IEEE 802.11b standard specifies that a 802.11b signal must be limited to a 22 MHz (-30 dB) double-sided RF bandwidth [51]. The minimum possible sampling rate, due to the Nyquist sampling criteria, is 44 MS/s. Assuming we quantize samples to 8 bits for each in-phase and quadrature sample, we would need 88 MB/s to transfer them through the USB link. Because this rate is too high for the current USB link, as introduced in Section I, two ways have been proposed to address the problem in the implementation of the 802.11b channel sounder; *Bandwidth Reduction Method* and *FPGA Matched Filtering Method*. In this section, we overview the two implementation options before providing more detail on the latter option, which is what we implement in this paper.

### A. Bandwidth Reduction Method

In the bandwidth reduction method implementation, the FPGA filters and downsamples the downconverted received signal. The filter is a cascaded integrator comb (CIC) filter with a 8 MHz RF bandwidth. Downsampling converts the sample rate from 64 MS/s to 16 MS/s. In short, the FPGA is used to reduce the data rate to something that can be transferred over a USB 2.0 link [52]. This implementation was developed to handle 802.11b demodulation using a PC for software-based reception [40]. It is possible to use the same sampled signal on the PC to estimate the channel impulse response on the PC as well; Figure 3(b) shows the block diagram of how such CIR estimation is performed in this implementation.

Because the sampling rate is significantly reduced, receiver performance is also reduced. The under-sampling effectively causes a significant reduction in the SNR, which reduces the ability of the 802.11b demodulator to make symbol decisions. Thus the receiver has a high bit error rate (BER) and packet error rate (PER) compared to a full-bandwidth receiver. In particular, packets transmitted at the 2 Mbps rate, modulated using DQPSK, are more sensitive to noise than packets sent at 1 Mbps (using DBPSK) and have very short range when using the bandwidth reduction method. The decreased bandwidth also reduces the resolution of CIR estimates, even when packets are received without error.

### B. FPGA Matched Filtering Method

Fundamentally, the rate limitations of a USB 2.0 link do not limit transfer of 802.11b signal information, since symbols are sent at 1 Msymbols/s [51]. DS-SS adds no information but causes the RF bandwidth to increase by a factor of 11. To reduce the received signal to samples at 1 MS/s, we must first despread on the USRP's FPGA. After despreading, symbol decisions can be made using only one samples per symbol. Further, as we will show, a subset of samples per symbol can be used as inputs for CIR estimation. In Figure 3 the block diagram of our proposed algorithm is compared to the bandwidth reduction implementation.

In this implementation, we first reduce the sampled data $r(t)$ to 32 MS/s. This received signal $r(t)$ was previously described in (4). Then, we despread, which involves convolution of the incoming signal with the known Barker code signal as given in (5). The output $q(t)$ still has a sample rate of 32 MS/s; however, it is not useful to send every sample to the PC. As a result, we will only send samples near the peaks in $q(t)$, as described in detail in Section V-C. The result is that the average data rate sent to the PC is within the rate limitations of a USB 2.0 link. The PC then performs the symbol detection and bit decoding operations as specified in the IEEE 802.11b standard. Our receiver implementation can consistently receive 802.11b packets sent at the 2 Mbps rate, and the reception range is up to 20 meters.

### *C. Matched Filtering on FPGA Implementation*

The challenge of implementing the matched filtering method on the FPGA is described in detail in this section. An FPGA has particular strengths and weaknesses; it is capable of performing highly parallel computations, and particularly adept at additions and multiplication by $-1$. However, it requires quite a bit of resources to implement general-valued multiplications. An FPGA is limited in the number of blocks and by its clock speed. Each storage or multiplication operation requires reserving blocks for that purpose. We describe in this section a computationally-efficient method to implement the 802.11b matched filter within the strict limitations of the given FPGA. Again, such computational simplifications will be used in real-world devices in order to optimize the tradeoff between cost and performance regardless of implementation in an FPGA or application-specific integrated circuit (ASIC).

There are three ways in which the implementation reduces computational complexity and data rate yet still provides a high-capability system implementation:

1) Integer multiplications and additions are rearranged in order to reduce the quantity of necessary multiplications,
2) Two memories are used to provide the ability to perform two operations per incoming sample, and
3) The peak of the matched filter output is determined and only samples near the peak are saved.

We detail these implementation decisions below.

*1) Multiplication Reduction:* Despreading is the convolution of the received signal with the PN code signal as described in previous section. In this implementation, the PN code signal $c(t)$ has duration 1 $\mu$s, which at 32 MS/s requires 32 samples. A direct implementation would require 32 multiplies and adds per sample. Figure 4 shows $c(t)$ and its sampled version, $c(iT_s)$. To avoid floating point multiplication, we quantize each $c(iT_s)$ to five bits, which we denote as $c_q(iT_s)$. Quantization to five bits is a tradeoff between resolution and multiplier space complexity.

Some values of $|c(iT_s)|$ are similar enough, as seen by inspecting sample values (•) in Figure 4, that when they are quantized to five bits, $|c_q(iT_s)| = |c_q(jT_s)|$, for some $j \neq i$. Since summation is much simpler than multiplication in an FPGA, it saves both time and complexity to first add (or subtract) samples with identical $|c_q|$ value, and then multiply the sum by its $|c_q|$ value. Using this rearrangement, we require 15 multiplications, instead of the 32 that would be required in the direct implementation. This can be expressed as,

$$\begin{aligned} q(nT_s) &= \sum_{i=0}^{31} c_q(iT_s) r((n-i)T_s) \\ &= \sum_{g=1}^{15} c_g \left[ \sum_{i \in S_g} \text{sgn}\left\{c_q(iT_s)\right\} r((n-i)T_s) \right] \end{aligned} \quad (15)$$

where $q(nT_s)$ is the $n$th sample of the match filter output $q(t)$, $S_g$ is the set of all indexes in the $g$th group, sgn$\{\cdot\}$ is the signum function, and $c_g$ is the multiplicative factor $c_q(iT_s)$ for all $i \in S_g$. The $S_g$ and $c_g$ for each group $g$ are listed in Table I.

*2) Two Memories:* An FPGA requires parallelization in order to complete the several additions and multiplications required at each new sampling time. Our implementation allows two clock cycles (clock rate of 64 MHz) per sampling time (sampling rate of 32 MS/s). During these two clock cycles, we must perform addition and multiplication as described above, and shift samples to allow space for the new incoming signal sample.

For this purpose, we use two 32-length arrays, which we refer to as mem and bmem. When a new sample is received, it is located at mem[0] while mem[1] to mem[31] are filled by bmem[0] to bmem[30]. In the next cycle, mem[0] to mem[31] are put in bmem[0] to bmem[31]. This process is depicted in Figure 5. As explained in previous paragraph, we first add the data in bmem, by group $g$, which is completed in one cycle. Then, multiplication by group multiplier $c_g$ is performed, and the results summed.

*3) Peak Selection:* The output of the FIR filter, $q(nT_s)$, has a 32 MS/s rate, and can not directly be fed to the USB 2.0 link. However, as described in Section IV, not all samples of $q(t)$ are necessary. With a sampling period of 31.25 ns, we will capture 344 ns of the signal within 11 samples, which means that 344 ns of the CIR can be estimated with only 11 samples per symbol. Typically, multipath excess delays for short range channels (such as WiFi links) will be limited to less than 344 ns, which corresponds to a 115 meter excess path length.

Eleven complex samples per symbol is as much as can be reliably carried by the USB link. The USB controller chip on the USRP board supports resolutions of $4, 8, 16$ bits per sample. At eight complex samples per symbol, the bit rate through the USB link is 176 Mb/s. This rate provides a reasonable rate for a USB 2.0 link as described at the beginning of this section.

The peak selection algorithm selects 11 out of each 32 samples per symbol as follows. First, the FPGA computes the power values $|q(nT_s)|^2$, $n = 1, \ldots, 32$. The index of samples with maximum power is denoted $n_{max} = \text{argmax}_n |q(nT_s)|^2$. The FPGA sends through the USB the samples from three samples before to seven samples after the peak power sample, *i.e.*, $\{q((n_{max} + i)T_s)\}_{i=-3}^{7}$.

Note that the proposed channel sounder finds samples near the maximum-power peak, not necessarily the line-of-sight (LOS) path. In a non-LOS dominant channel, if the LOS path arrives within three samples prior to the maximum-power peak, the proposed system records the full CIR.

## VI. Experimental Results

In all cases, we load an Ettus Research USRP (rev 4.5) with the code described in Section IV-E. The RF front end is a RFX2400 daughterboard (rev 30), also from Ettus Research. The antenna is a 2400-2480 MHz sleeve dipole antenna with omnidirectional pattern in the horizontal plane and a 3 dBi gain. The total cost of the hardware is $975. The USRP is connected to a Dell Inspiron laptop running Python and Matlab. The Python (GNU radio) code collects data from the USB, demodulates the packet data, and writes to a file. The Matlab code performs the averaging required in (8) and then displays and stores the impulse response estimate $\hat{h}(t)$. From the stored $\hat{h}(t)$, the deconvolution described in Section IV-E is performed in post-processing.

### *A. Demodulator*

In this section, we measure the packet reception rate of the implemented 802.11b channel sounder system. As described earlier, it is important to both estimate channel impulse response and estimate packet data. Since the symbol sequence is required to estimate $\hat{h}(t)$ in (8), we do not proceed with CIR estimation when packet data does not pass the CRC test. Equally important, the MAC address of a transmitter is included in the packet header, and this is necessary to distinguish packets originating from different transmitters.

We compare the packet reception rate between the two implementions described in Section V, the bandwidth reduction method (the prior work), and the FPGA matched filtering method (the proposed system implementation). We expect that the bandwidth reduction method has lower SNR and will have lower packet reception rates for the identical setup.

In the first experimental scenario, we examine packet reception rate in the absence of interference. We configure a test transmitter, a D-Link 802.11b wireless router (model DI-614+), to broadcast a beacon packet at a basic rate (1 or 2 Mbps) every 200 ms (5 packets/sec). The router and receiver are placed in a shielded anechoic chamber and separated by 6.0 meters. The packet reception rate is recorded four times for three minutes. The number of received packets for the two system implementations is presented in Table II. The results show that the FPGA matched filtering method outperforms the bandwidth reduction method by successfully demodulating 1.6 times more packets.

### *B. Channel Measurement*

In this section we first perform two experimental validations on our implementation using known channels between the transmitter and the receiver. Then, we perform an experimental measurement campaign to measure a large number of CIRs in outdoor areas in and around Salt Lake City, Utah. We provide measurement results and summarize the measured delay characteristics.

*1) Validation:* To validate the CIR estimation system, we create a channel with known impulse response out of RF hardware and cable. The antenna is removed from the router (described previously) and the receiver and both are connected to a coaxial cable. The results are compared to the output of a network analyzer and it agrees with the presented channel sounder. We create two different known channels:

1) Single-path: The transmitter is connected to an attenuator, whose output is connected to the receiver.
2) Double-path: The transmitter cable is connected to a RF splitter with two outputs, one connected to a short cable, and another to a long cable. Both paths contain an attenuator. The outputs of the cables are input to a signal combiner and then connected to the receiver. Figure 6 shows a schematic of the double-path channel simulator.

*Single-path:* With the transmitter and receiver set up in the single-path channel, we record several measured CIR estimates $\hat{h}_1(t)$. Solid lines in the top plot of Figure 7-(a) shows five different measured CIR estimates $\hat{h}_1(t)$. In that figure, the known CIR for a single-path channel, which is simply $R_s(t)$, is plotted in dashed line for comparison. Since $\hat{h}_1(t)$ is almost the same as a single pulse $R_s(t)$, it is apparent that the channel has only one path, *i.e.* $L = 1$. Figure 7-(b) presents the result of deconvolving $\hat{h}(t)$ as given in (14).

*Double-path:* In the next experiment, the router is connected to the receiver via two cables, one with length 0.3 m and a second with length of 29 m, attenuators, and RF splitter and combiner. We first measure the impulse response using a vector network analyzer, from which we find that the difference in delay between the two paths is 140 ns. The amplitude difference between the two paths is measured to be 9.0 dB by the following procedure. We disconnect the combiner (seen in Figure 6) and use a LadyBug power sensor (LB479A) to measure the power at the two cables that feed the combiner. The power in the longer cable is 9.0 dB lower than the power in the shorter cable.

Figure 7(a), presents $\hat{h}_2(t)$, the CIR estimate of double-path channel. Since $\hat{h}_2(t)$ is wider in time than a single pulse $R_s(t)$, dashed line in that figure, it is apparent that the channel has more than one path, *i.e.* $L > 1$. The deconvolution algorithm of (14) is applied and the resulting estimate in plotted in Figure 7-(c). The results show that the channel has two paths, the later path arrives 150 ns after the first path and with 9 dB less received power.

In Figure 7(a) and (b), five different measurements (taken in different times) are almost the same for each validation scenario which represents the repeatability of the channel sounder. Observation of Figures 7(b) and 7(c) lead us to the conclusion that the noise floor for the channel sounder is approximately 20 dB below the signal power. Since the PN coding gain of the Barker code is $20 \log_{10} 11 \approx 20.8$ dB, the observed self-interference floor is expected.

### C. Drive-Test CIR Measurement Campaign

We use our channel sounder to measure channel impulse responses in three residential, two commercial areas, and one downtown area in Salt Lake City. The residential areas are comprised of one to three story single-family homes and apartment buildings. The commercial areas include streets near strip malls, low-rise office buildings, and heavy vehicle traffic. The downtown area is an urban canyon of high-rise office buildings on both sides of the streets. In each area, the receiver antenna is on the outside of a vehicle that drives at typical speeds on city streets. Approximately five minutes of measurements are recorded in each area. In the course of six five-minute drive-test measurements, a total of three million CIR measurements are recorded. Figure 8 presents a typical example of the channel estimate $\hat{\boldsymbol{\alpha}}$ from each area.

In order to compare different multipath channels and to develop some general design guidelines for wireless systems, parameters that grossly quantify the multipath channel are used. The time dispersive properties of wide band multipath channels are most commonly quantified by their mean excess delay $\bar{\tau}$ and RMS delay spread $\sigma_\tau$, which are defined as follows [24]:

$$\bar{\tau} = \frac{\sum_k |\alpha_k|^2 \tau_k}{\sum_k |\alpha_k|^2}, \quad \sigma_\tau = \sqrt{\bar{\tau^2} - (\bar{\tau})^2}, \text{ where } \bar{\tau^2} = \frac{\sum_k |\alpha_k|^2 \tau_k^2}{\sum_k |\alpha_k|^2} \tag{16}$$

Where $\alpha_k$ and $\tau_k$ are respectively complex amplitude gain and excess delay of $k$th path. In Table III the average mean excess delay, average and maximum RMS delay of channel responses in these area are presented.

The drive test experiment demonstrates that the channel measurement system performs as expected. The CIR estimates in Figure 8 and the mean delay and RMS delay spread results in Table III show characteristics that are intuitively appealing. In residential areas, there are few multipath at significant delays, while in commercial and in downtown areas, there are multipath at longer delays, and with amplitudes close to the amplitude of the first arriving path. Average RMS delay spreads, which are averages over all measured channels (around six million measurements), range from 12.1 ns in the lowest-multipath (residential), to 44.6 ns in the highest-multipath (downtown) area.

## VII. Conclusion

Future wireless networks are envisioned that rely on the real-time estimation of CIR from received WiFi packets for the purposes of cross-layer security, localization, and environmental imaging. We present a CIR estimation system using an inexpensive and open source hardware and software platform to enable these emerging areas of research. We show how accurate CIR estimation can be performed using a resource-constrained FPGA, which provides a proof-of-concept for future commercial devices. A deconvolution algorithm is implemented to refine CIR estimates, and measurements using known channels are conducted to verify estimator accuracy. A drive test is performed, recording CIR in six different areas in a city, as an example of the use of the CIR estimation system.

As an open source platform, others may improve upon the work described here. As of the submission date, our implementation and source code have been downloaded at least 430 times since its first posting in March, 2008, not counting those downloads from hosts other than our own. Work remains to additionally implement 802.11 transmission, in addition to reception and CIR estimation, on future devices, for example, on the USRP 2.0. In general, the additional capability of the USRP 2.0 allows implementation of other transceiver features in addition to the CIR estimation described here. Future work will also address MIMO CIR estimation, using a bank of synchronized software radios. Low complexity MIMO channel sounding will likely benefit the development of future cross-layer techniques in multiple-antenna wireless networks.

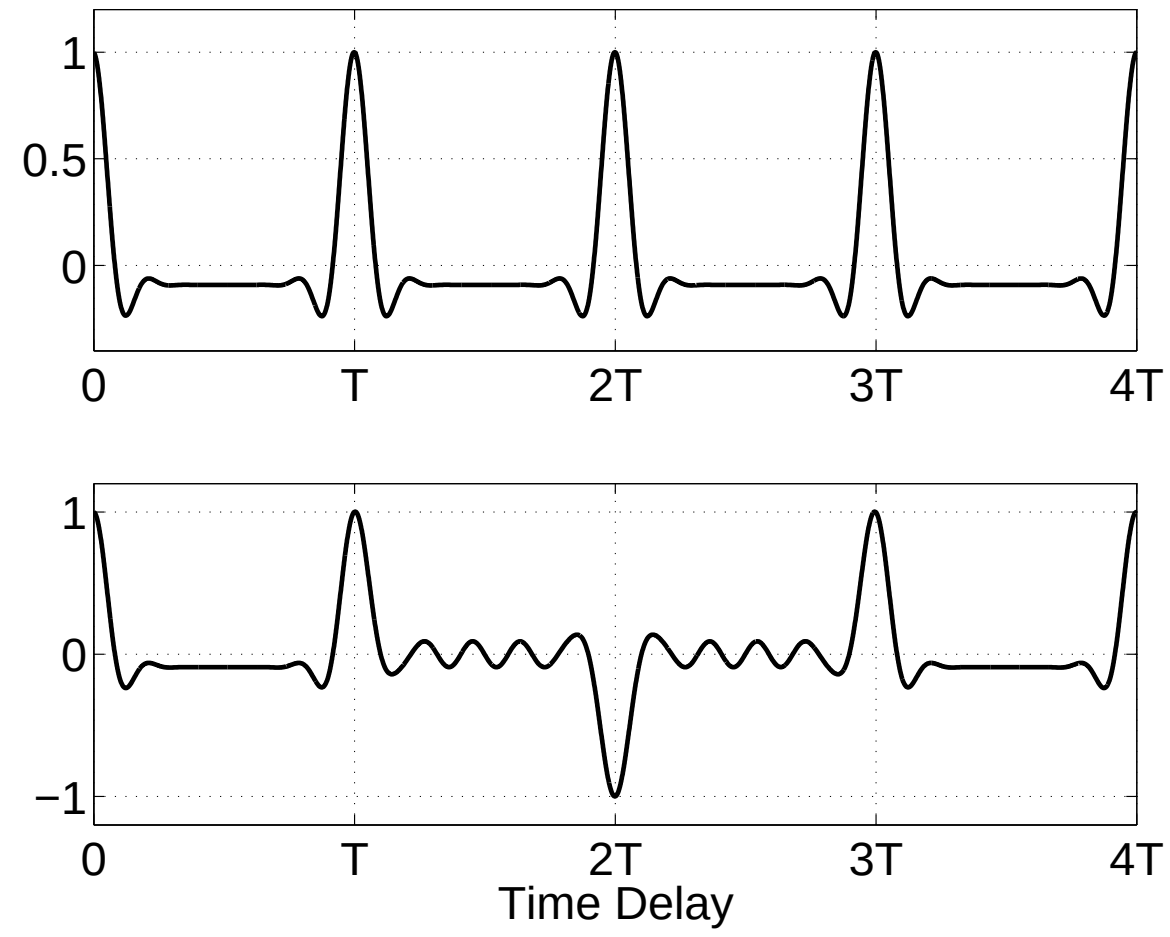

Fig. 1. Correlation output signal $q(t)$ in one-path channel ($L = 1$ and $\alpha_0 = 1$) when (Top) receiving an unmodulated signal (*i.e.*, $\mathbf{b} = [1, 1, 1, 1, 1]$) (Bottom) receiving a signal modulated with $\mathbf{b} = [1, 1, -1, 1, 1]$.

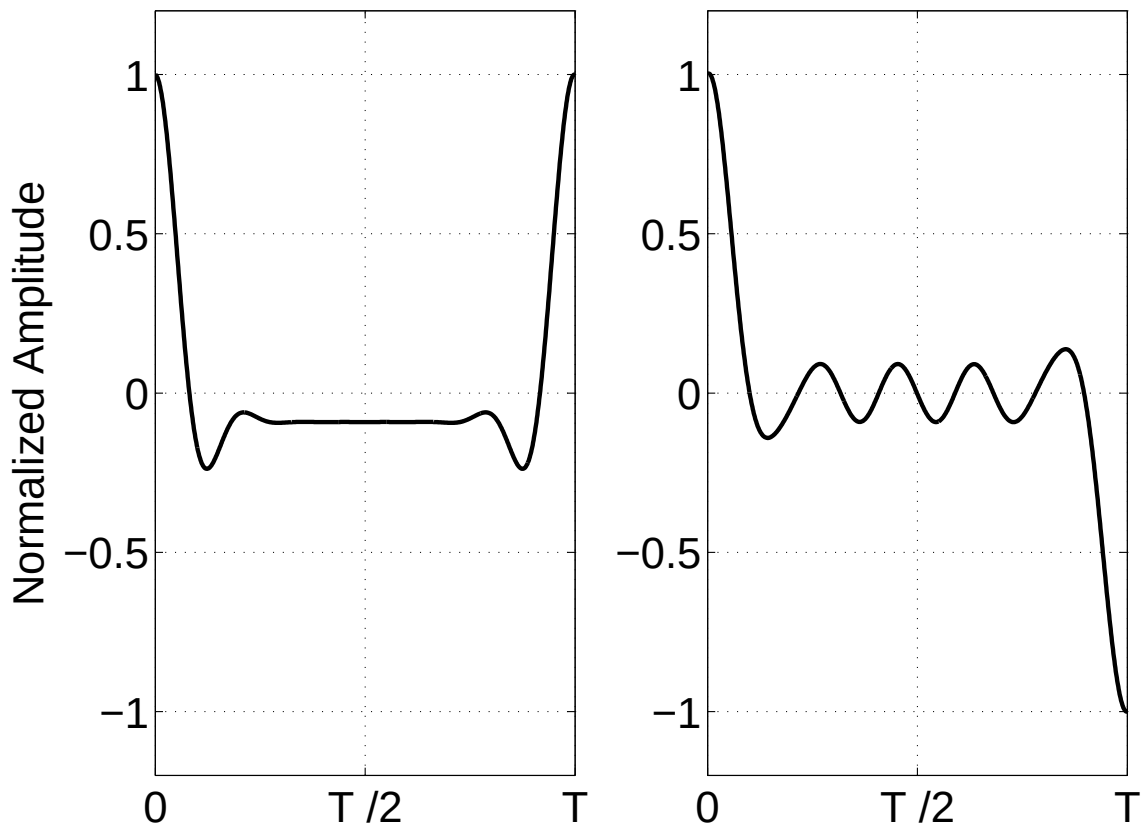

Fig. 2. Normalized symbol-period length correlation functions (Left) $R_s(t)$ and (Right) $R_o(t)$, both given in (7).

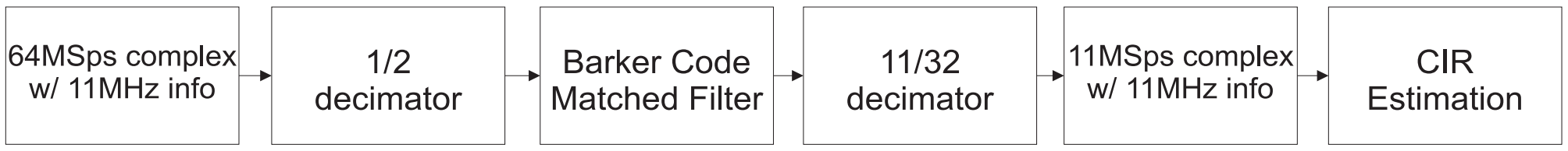

(a)

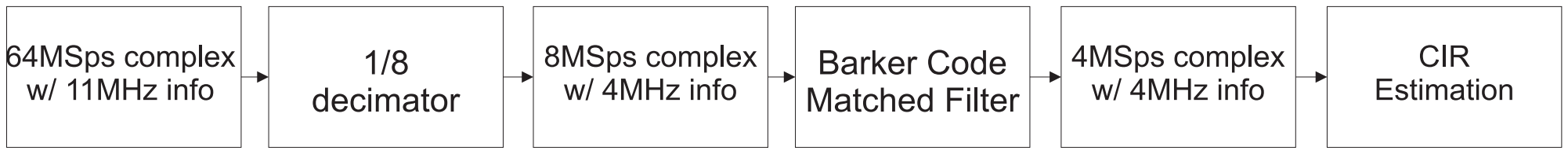

(b)

Fig. 3. Block diagram of (a) FPGA matched filtering method, and (b) bandwidth reduction method.

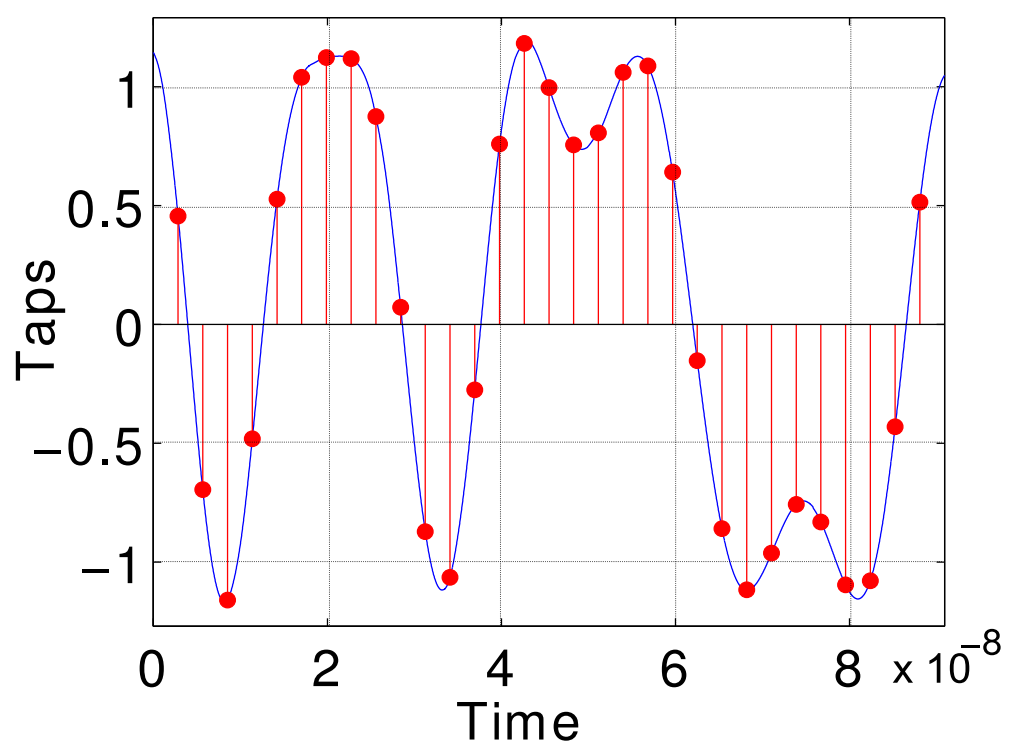

Fig. 4. Samples (•) of PN code signal $c(t)$, or equivalently, taps of the matched filter.

| $g$ | Multiplier $c_g$ | Index Set $S_g$ |
|---|---|---|
| 1 | 19 | $\{16, 28\}$ |
| 2 | 18 | $\{3, 7, 23, 24, 31\}$ |
| 3 | 17 | $\{11, 12, 19, 22, 15\}$ |
| 4 | 16 | $\{25\}$ |
| 5 | 15 | $\{6\}$ |
| 6 | 14 | $\{8, 20, 22\}$ |
| 7 | 13 | $\{4, 13\}$ |
| 8 | 12 | $\{5, 14, 17\}$ |
| 9 | 11 | $\{29\}$ |
| 10 | 10 | $\{10\}$ |
| 11 | 8 | $\{0, 26, 27\}$ |
| 12 | 7 | $\{1, 30\}$ |
| 13 | 4 | $\{18\}$ |
| 14 | 2 | $\{9\}$ |
| 15 | 1 | $\{21\}$ |

TABLE I

INDICES BY GROUP $g$ AND THE GROUP'S MULTIPLIER VALUE $c_g$.

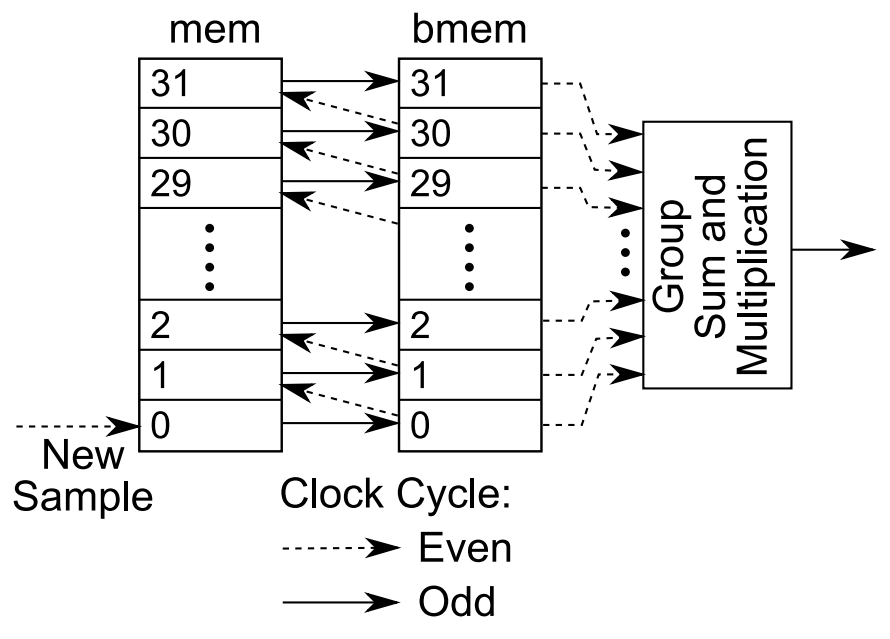

Fig. 5. Memories mem and bmem are used to accept a new sample, and shift data, in two cycles, to allow for summation and multiplication.

| | Bandwidth Reduction | FPGA Matched Filtering | Ratio |
|---|---|---|---|
| 1 | 462 | 718 | 1.55 |
| 2 | 476 | 712 | 1.5 |
| 3 | 442 | 736 | 1.67 |
| 4 | 438 | 730 | 1.67 |

TABLE II

RECEIVED PACKET COUNT FOR TWO DEMODULATION METHODS

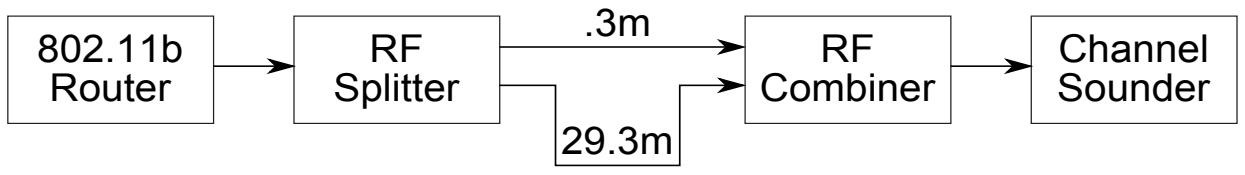

Fig. 6. Two-path channel experimental setup.

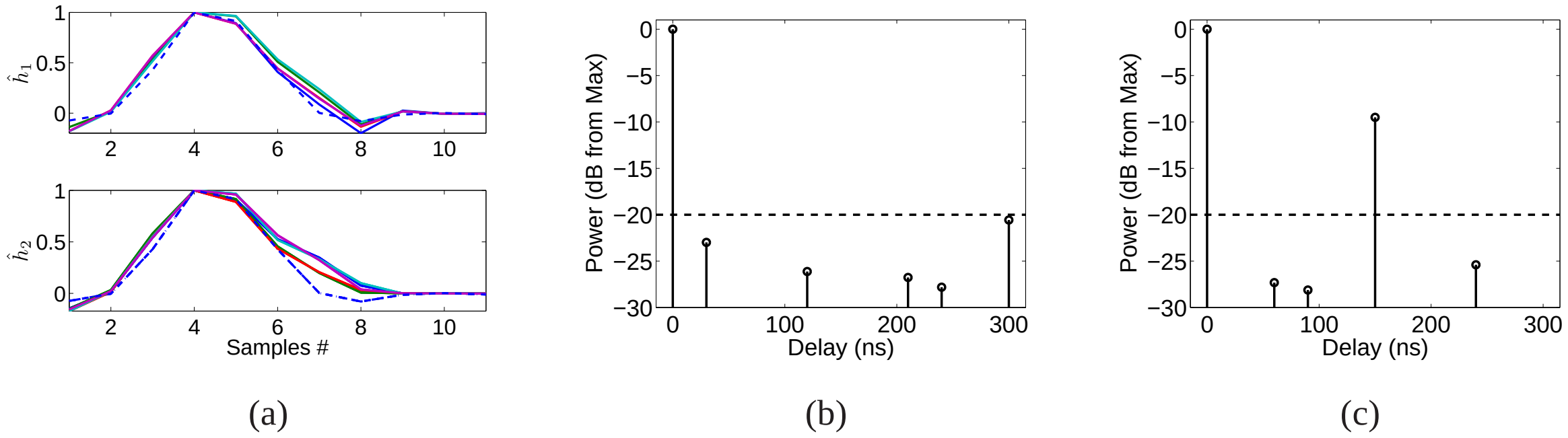

Fig. 7. Single- and double-path results: (a) $\hat{h}$ for single-path (Top) and double-path (Bottom), both showing five different measurements (solid lines) and ideal CIR $R_s[n]$ (dashed line); Deconvolved $\hat{\boldsymbol{\alpha}}$ for (b) single-path; and (c) double-path. In (b) and (c), dashed line presents the noise floor.

| | Res. 1 | Res. 2 | Res. 3 | Com. 1 | Com. 2 | Downtown |
|---|---|---|---|---|---|---|
| Average $\bar{\tau}$ (ns) | $5.27$ | $6.99$ | $8.18$ | $41.05$ | $37.84$ | $49.68$ |
| Average $\sigma_\tau$ (ns) | $11.6$ | $23.27$ | $10.79$ | $58.54$ | $52.47$ | $68.9$ |
| Max. $\sigma_\tau$ (ns) | $41.83$ | $53.86$ | $21.98$ | $103.43$ | $93.74$ | $127.11$ |

TABLE III

RMS DELAY SPREAD AND MEAN EXCESS DELAY STATISTICS FOR MEASURED AREAS.

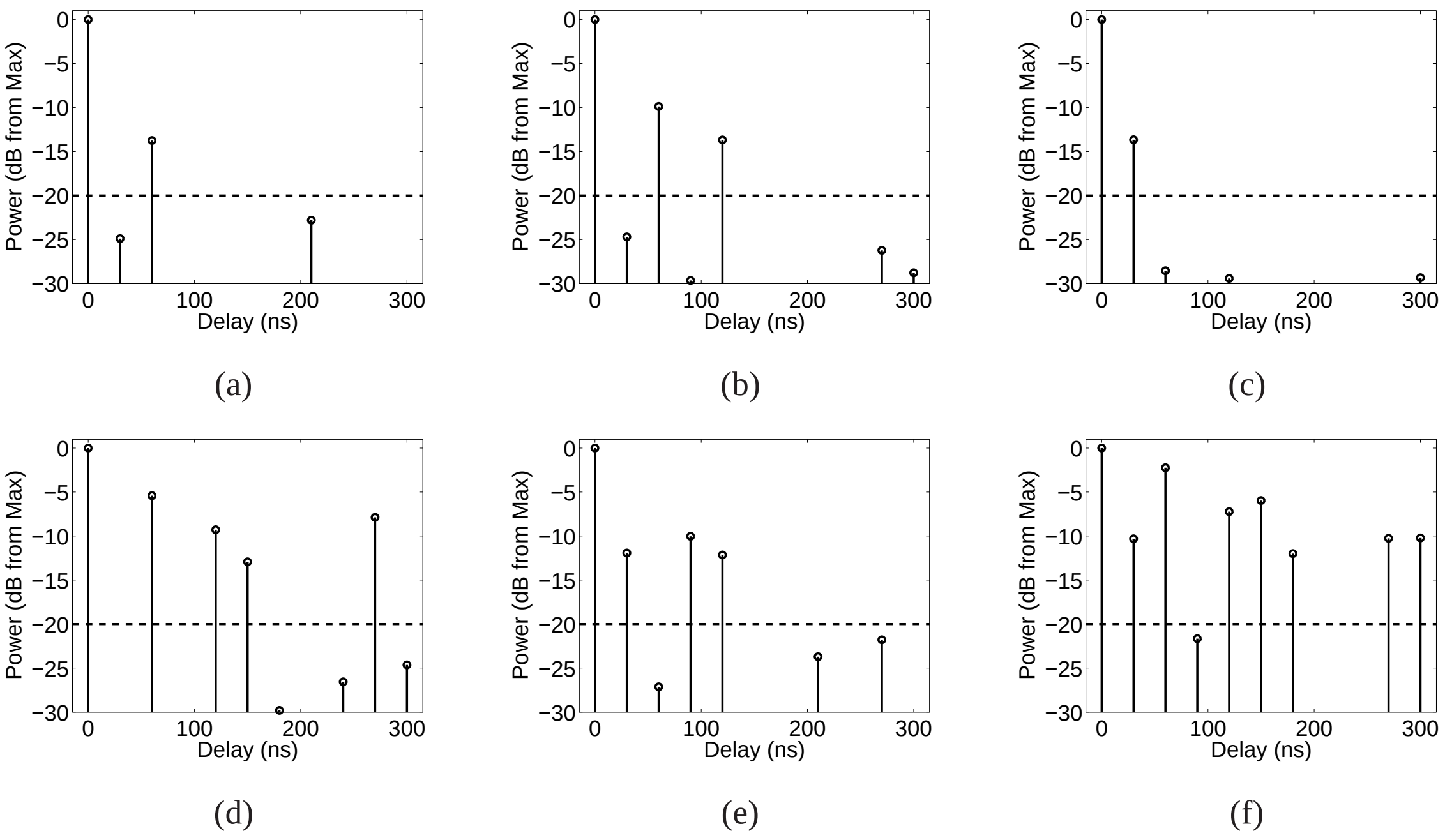

Fig. 8. Typical CIR, $\hat{\alpha}$, for the six different measured environments in Salt Lake City: (a)-(c) residential areas, (d)-(e) commercial areas, and (f) downtown. In these figures dashed line presents the noise floor.